\documentclass[twocolumn,pre,showpacs]{revtex4}

\usepackage{graphicx}
\usepackage{dcolumn}
\usepackage{bm}

\begin{document}

\title{A theory of growth by differential sedimentation,\\
 with application to snowflake formation}

\author{C.D. Westbrook}

\author{R.C. Ball}

\affiliation{Department of Physics, University of Warwick, Coventry, UK. }

\author{P.R. Field}

\affiliation{Met Office, Exeter, UK. }

\author{A.J. Heymsfield}

\affiliation{National Centre for Atmospheric Research, Boulder, Colarado, USA. }

\date{\today{}}

\begin{abstract}
A simple model of irreversible aggregation under differential sedimentation
of particles in a fluid is presented. The structure of the aggregates
produced by this process is found to feed back on the dynamics in
such a way as to stabilise both the exponents controlling the growth
rate, and the fractal dimension of the clusters produced at readily
predictable values. The aggregation of ice crystals to form snowflakes
is considered as a potential application of the model. 
\end{abstract}


\maketitle

\section{Introduction}

Simple models of cluster-cluster aggregation have been the focus of
a great deal of interest, particularly over the last two decades.
The structure of aggregates formed through a variety of dominating
mechanisms (eg. diffusion limited \cite{meakin}, reaction limited
\cite{balletal} and ballistic motion \cite{jullienandkolb}) have
been studied through theoretical, experimental, and computational
work.

Another aggregation mechanism which is relevant to several physical
systems is that of differential sedimentation. Particles with a range
of size and/or shape will almost inevitably sediment through a fluid
at different speeds under the influence of gravity, leading to collisions.
If there is some mechanism by which the particles stick on contact
then aggregates will be formed. An example of this kind of phenomenon
is the aggregation of ice crystals in Cirrus clouds. Small `pristine'
ice particles are formed at the top of the cloud, and proceed to fall
through it, colliding with one another and sticking to produce aggregates
(snowflakes).

The aim of this paper is to provide a simple model for growth by differential
sedimentation which captures the essential physics of the system in
the inertial flow regime, and to consider its application to snowflake
formation. It is divided into five main parts - a description of the
model and the assumptions underlying it; details of computer simulations
and the results obtained from them; a theory section which offers
an argument to account for the behaviour observed in the simulations;
an investigation of the model's applicability to snowflake formation;
and a concluding discussion.  The simulation results and their comparison to real cloud data have been presented briefly in a separate letter \cite{ourshortpreprint}.

\section{Model}

We focus on the dilute limit, where the mean free path between cluster-cluster
collisions is large compared to the nearest neighbour distance between
clusters. In this regime we can limit our interest to individual binary
collision events, ignoring spatial correlation. As further simplifying
approximations, we assume that clusters have random orientations which
do not significantly change during a close encounter, that collsion trajectories are undeflected by hydrodynamic interaction, and that any
cluster-cluster contacts result in a permanent and rigid junction.

\begin{figure}
\includegraphics{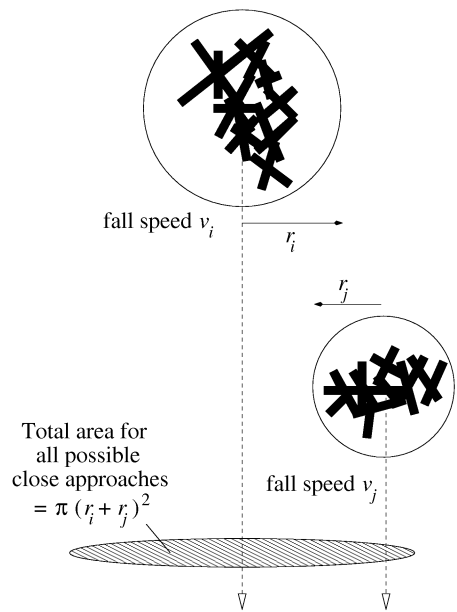}
\caption{\label{illustration}Illustration showing a possible scenario in which the centres of a pair of clusters falling at a relative speed $|v_i-v_j|$ come within a distance $(r_i+r_j)$ of one another (a close approach). The shaded circle illustrates the total area encompassing all possible close approach trajectories $=\pi(r_i+r_j)^2$.}
\end{figure}

In order to sample the collisions between clusters, we first formulate
a rate of close approach. For any two clusters $i$, $j$ with nominal
radii (see below) $r_{i}$ and $r_{j}$ respectively and fall speeds
$v_{i}$ , $v_{j}$ , the frequency with which their centres pass
closer than a distance $(r_{i}+r_{j})$ is proportional to
the total area over which trajectories yielding a close approach event are possible, and
the relative speed of the pair. This is illustrated in figure \ref{illustration}. The rate constant for approach closer than centre-to-centre separation $(r_{i}+r_{j})$ is therefore given by:
\begin{equation}
\Gamma _{ij}=\pi (r_{i}+r_{j})^{2}\left|v_{i}-v_{j}\right|.\label{gamma}\end{equation}
 In our computer simulations the nominal radii are chosen to fully
enclose each cluster and the close approach rate calculated above is exploited to preselect candidate collision events.  Collisions are accurately sampled by 
sequentially choosing pairs of clusters with probability proportional to $\Gamma _{ij}$, checking each pair for collision along one randomly sampled close approach
trajectory, and correspondingly joining that cluster pair if they do indeed collide. In the theoretical arguments presented in section four,
we make the simplifying assumption that all close approaches lead
to collisions (or at least a fixed fraction of them do), using nominal
radii based on fractal scaling from the cluster masses.

The model is completed by an explicit form for the fall speeds entering
equation (\ref{gamma}). We assume that the clusters are at most only partially penetrated by the fluid flow past them, so that cluster radius is the relevant length governing the drag force law. Then qualitatively and by dimensional argument we expect the same drag behaviour as for a falling sphere, which may be written in the form:
\begin{equation}
F_{d}=\rho \nu _{k}f(R_e)\label{drag}\end{equation}
 where $f$ is a function of the Reynolds number $R_e=rv/\nu _{k}$
alone, $\rho $ is the density of the surrounding fluid, and $\nu _{k}$
is the kinematic viscosity. Although details of the function $f(R_e)$ should be different from spheres, we still expect to have inertial and Stokes regimes where $f$
takes the forms:
\begin{eqnarray}
\label{flow}
f(R_e)\sim
\left\{
\begin{array}{c}
R_e^2\text{ for inertial flow}\\
R_e\text{ for viscous flow}
\end{array}\right\}\;.
\end{eqnarray}
We consider below the general form $f(R_e)\sim R_e^{1/\alpha }$, with
$\alpha $ as an adjustable parameter in order gain understanding spanning the two extreme regimes. Setting the drag force equal
to the weight $mg$ of the cluster, the terminal velocity is then
given by\begin{equation}
v\sim \frac{\nu _{k}}{r}\left(\frac{mg}{\rho \nu _{k}^{2}}\right)^{\alpha }\label{fallspeed}\end{equation}
where $\alpha =\frac{1}{2}$ for inertial flow and $\alpha \textrm{=1}$
for viscous flow.  A more complete discussion of the fall speed is given by Mitchell \cite{mitchell96}, but provided $d_f \ge 2$ so that cluster projected area scales as the square of cluster radius, this reduces to a simple crossover between the above limits.  The empirical crossover is very slow, spread over some three decades of Reynolds number, so fixed intermediate values of $\alpha$ can reasonably approximate behaviour over a significant range \cite{mitchell96}. In our simulations we took the radius determining
the fall velocity to be proportional to the radius of gyration, and in
our theoretical calculations we simply used the same nominal radii
as for the collision cross sections above.

\section{Computer simulations and results}

The primary particles used at the beginning of the simulations were
rods of zero thickness, half of which had a length (and mass) of unity,
and half of which were twice as long and massive. Purely monodisperse
initial conditions are not possible in this model, since $|v_{i}-v_{j}|$
would be zero. Apart from this special case however, it is anticipated
that the asymptotic behaviour of the system should be insensitive
to the initial distribution, and indeed the results described in this
section appear to be preserved for a variety of starting conditions.

\begin{figure*}
\includegraphics{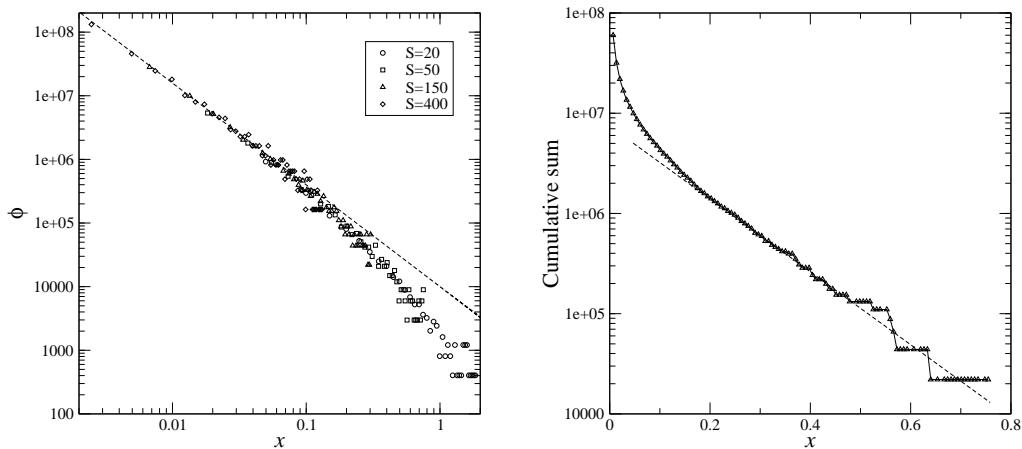}
\caption{\label{distribution}Scaling of the cluster mass distribution. The
left panel shows how the rescaled cluster size distribution $\phi =s(t)^{2}n_{m}(t)$
converges to a universal function of rescaled cluster size $x=m/s(t)$,
where the data are overlayed for different values of the weight average
cluster size, $s(t)=20,50,150,400$. The scales are logarithmic and
a least squared fit $\phi(x)\sim x^{-1.6}$ for $x\leq 10^{-2}$
is shown by the dashed line. In the right hand panel $\int _{x}^{\infty }\phi (x')\text {d}x'$
is shown on a semi-log plot, illustrating the exponential tail (dashed
line is intended to guide the eye). Both simulations began with 250,000
rods, and used $\alpha =0.55$ in the sedimentation law.}
\end{figure*}

In aggregation models it is typically the case (eg. Vicsek and Family
\cite{vicsekandfamily}) that after the distribution has had time
to `forget' its initial conditions it will approach a universal shape.
This is usually expressed by the `dynamical scaling' ansatz, which
states that as $m,s\rightarrow \infty$:
\begin{equation}
n_{m}(t)=s(t)^{-2}\phi\left[\frac{m}{s(t)}\right]\label{dynamicalscaling}
\end{equation}
 where $n_{m}(t)$ is the number of clusters of mass $m$ at time
$t$, and the rescaled distribution $\phi $ is a function of $x=m/s(t)$
alone. The quantity $s(t)$ is a characteristic cluster mass, and
for non-gelling systems one expects that a suitable choice is given
by the weight average cluster mass, $s(t)=\sum _{i}m_{i}^{2}/\sum _{i}m_{i}$.
Using this choice our simulation data conform well to scaling, as
shown in the left panel of figure \ref{distribution}.

The shape of the rescaled distribution was studied. A plot of $\int _{x}^{\infty }\phi (x')\text {d}x'$
as a function of $x$ is shown in the right panel of figure \ref{distribution}
and shows an exponential decay for very large $x$, with a `super-exponential'
behaviour taking over as $x$ approaches unity from above. This behaviour
appears to be universal for all values of $\alpha $ in the range
studied.

For $x\ll 1$ the qualitative form of $\phi(x)$ was found to fall
into two distinct catagories depending on the value of $\alpha $.
For $\alpha \geq \frac{1}{2}$ the distribution appears to diverge
as a power law: $\phi(x\rightarrow 0)\sim x^{-\tau }$, as shown
in figure \ref{distribution} for $\alpha =0.55$. The exponent $\tau $
was found to be approximately constant at $\tau \simeq 1.6\pm 0.1$
over the range $\frac{1}{2}\leq \alpha \leq \frac{2}{3}$. For $\alpha <\frac{1}{2}$
the distribution was found to be peaked, with a maximum at some small
size $x_{m}$, followed by a power law decay for $x_{m}\ll x\ll 1$.

\begin{figure*}
\includegraphics{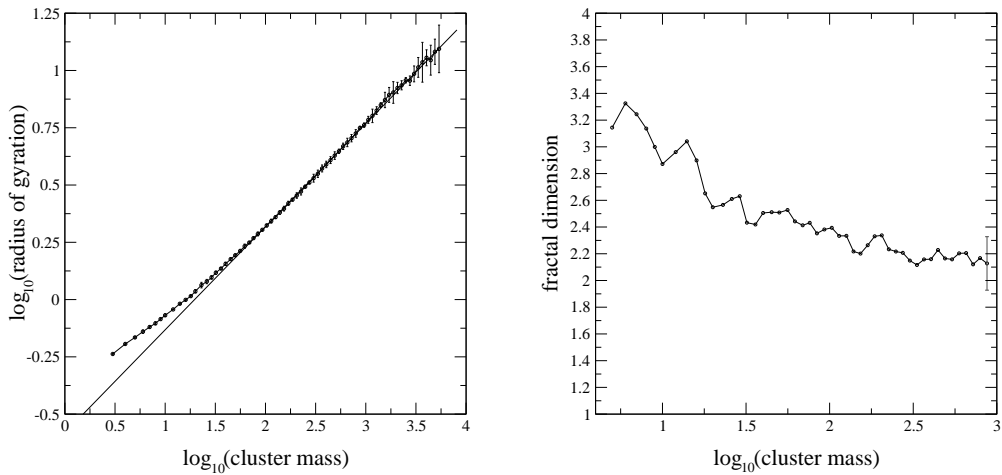}
\caption{\label{rgm}Left hand panel shows a log plot of radius of gyration as a function
of cluster mass for $\alpha =0.55$, averaged over four runs of 250,000
initial rods. Solid line indicates the theoretical prediction for
the fractal dimension. The right hand panel shows the inferred fractal
dimension as a function of cluster mass. Error bars are one standard
deviation. Data points with $\sigma >0.3$ have not been plotted.}
\end{figure*}

\begin{figure}
\includegraphics{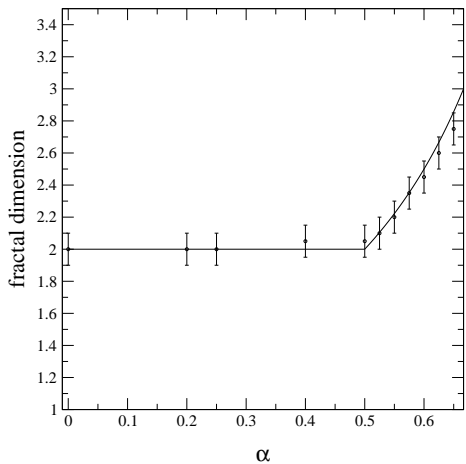}
\caption{\label{dfvsalpha}Variation of the fractal dimension as a function
of the parameter $\alpha $. Circles are simulation data, solid line
indicates theoretical prediction.}
\end{figure}

Comparison with other aggregation models suggests that the clusters
produced are likely to be fractal in their geometry, and in particular
cluster mass and (average) radius should be in a power law relationship
$m\sim r^{d_{f}}$ where $d_{f}$ is the fractal dimension. A log
plot of radius of gyration against mass for all the clusters produced
over the course of the simulation is shown in figure \ref{rgm}. Also
shown in this figure is the logarithmic derivative of the above plot,
which shows the variation in the apparent fractal dimension of the
clusters with size. From this plot, it seems that the fractal dimension
approaches an asymptotic value as $m\rightarrow \infty $; in the
case shown ($\alpha =0.55$) we estimate this value as $d_{f}\simeq 2.2\pm 0.1$.
The value of this limiting fractal dimension was found to vary
with $\alpha $ as shown in figure \ref{dfvsalpha}.  Note that our assumption $d_f \ge 2$, required to support the assumed fall speed relationship, is indeed satisfied for the physical range $\alpha \ge 1/2$.

\section{Theory}

The most common theory used to describe cluster-cluster aggregation
problems is that of von Smoluchowksi \cite{smoluchowski}, which provides
a set of mean-field rate equations for the evolution of the cluster
mass distribution: \begin{equation}
\frac{\text {d}n_{k}(t)}{\text {d}t}=\frac{1}{2}\sum _{i+j=k}K_{ij}n_{i}(t)n_{j}(t)-n_{k}(t)\sum _{j=1}^{\infty }K_{ij}n_{j}(t).\label{smol2}\end{equation}
 where $n_{k}(t)$ is the number of clusters of mass $k$ at time
$t$ (per unit volume). The kernel $K_{ij}$ contains the physics
of the problem, being a symmetric matrix, the elements of which govern
the rate of aggregation between pairs of clusters expressed (only)
in terms of their masses $i$ and $j$. Analytical solutions of Smoluchowski's
equations have not been obtained except for a few special cases of
$K_{ij}$. However, Van Dongen and Ernst \cite{vdande85} have shown
that for non-gelling kernels (see below) the solutions approach the
dynamical scaling form of equation (\ref{dynamicalscaling}) in the
large-mass, large-time limit; substituting this into equations (\ref{smol2})
allows one to obtain some information about the asymptotic behaviour
of the rescaled cluster size distribution $\phi(x)$.

To apply this theory we need to compute the reaction rates $K_{ij}$
, which means averaging collision rate with respect to cluster geometry
at fixed masses. This we estimate by substituting averages from fractal
scaling for the radii in equations (\ref{gamma}) for the close approach
rate and (\ref{fallspeed}) for the fall speeds, and assuming constant
collision efficiency leading to: \begin{equation}
K_{ij}\sim \left|i^{\alpha -1/d_{f}}-j^{\alpha -1/d_{f}}\right|\left(i^{1/d_{f}}+j^{1/d_{f}}\right)^{2}.\label{kernel3}\end{equation}
 Van Dongen and Ernst's analysis is sensitive to two exponents
characterising the scaling of the coagulation kernel in the limit
$1\ll i\ll j$,\begin{equation}
K_{ij}\sim i^{\mu }j^{\nu }\end{equation}
which in our case yields: \begin{eqnarray}
\mu  & = & \min (0,\alpha -d_{f}^{-1})\label{exponentmu}\\
\nu  & = & \max (\alpha +d_{f}^{-1},2d_{f}^{-1}).\label{exponentnu}
\end{eqnarray}
 A third exponent combination $\lambda =\mu +\nu =\alpha +d_{f}^{-1}$ controls
the growth of the average cluster mass through the differential equation
$\dot{s}(t)=ws(t)^{\lambda }$, where $w$ is a constant, and for
the non-gelling case we require $\lambda \leq 1$.

Our identification of the exponent $\nu $ is crucial to a mechanism
by which the fractal dimension can control the dynamics. If the fractal
dimension is low enough, then the exponent $\nu $ will exceed unity.
However, Van Dongen \cite{vandongen} has shown that the Smoluchowski
equations predict the formation of an infinite cluster instantly in
such a situation. In a finite system this clearly cannot occur, and
it simply means that a few clusters will quickly become much larger
than the others with their growth dominated by accretion of small
ones. In this scenario the growth of the large clusters approaches
that of ballistic particle-cluster aggregation, where it has been
shown by Ball and Witten \cite{ballandwitten} that the fractal dimension
of the clusters produced tends to $d_{f}=3$. This increased fractal
dimension reduces the value of $\nu $, forcing it back to a value
of one if $\alpha \leq \frac{2}{3}$. Through this feedback mechanism,
a bound is placed on the fractal dimension $d_{f}\geq \max [2,(1-\alpha)^{-1}]$
for $\alpha \leq \frac{2}{3}$.

The system could perhaps settle in a state where $\nu <1$. However,
the growth in such a regime is much less biased towards collisions
between clusters of disparate sizes, and the distribution is relatively
monodisperse. This would tend to make collisions between clusters
of a similar size likely, leading to much more open structures, with
a lower fractal dimension, in turn acting as a feedback mechanism
to increase the value of $\nu $. The authors suggest that, at least
over some range of $\alpha $, this effect will force the system towards
the $\nu =1$ state. The discontinuity in the polydispersity of the
system at $\nu =1$ forces the system to organise itself such that
it can remain at that point. This is similar to the argument put forward
by Ball \textit{et al} \cite{balletal} for reaction limited aggregation.

If it is accepted that $\nu \rightarrow 1$ then the fractal dimension
of the clusters produced ought to be directly predictable from equation
(\ref{exponentnu}) : \begin{equation}
d_{f}=\max [2,(1-\alpha )^{-1}],\quad \alpha \leq \frac{2}{3}.\label{predictdf}\end{equation}
 A curve showing this theoretical behaviour is superimposed on the
simulation data in figure \ref{dfvsalpha}, and appears to show good
agreement up to $\alpha \simeq \frac{2}{3}$. For $\alpha>\frac{2}{3}$
the theoretical prediction is that $d_{f}=3$ and $\nu =\alpha +\frac{1}{3}>1$,
but because of its somewhat pathological nature we have not attempted
to make simulations in this regime. It is however clear from the extrapolation
of our results in figure \ref{dfvsalpha} that this is likely to hold.

Obtaining an exact form for the cluster size distribution $\phi (x)$
is a non-trivial exercise. However, following the methodology of Van
Dongen and Ernst \cite{vdande85}, we consider the small-$x$ behaviour
of $\phi (x)$ when $d_{f}<\alpha ^{-1}$ (ie. $\mu <0$). In such
a regime the small-$x$ behaviour is dominated by collisions between
clusters of disparate sizes; the gain term in the Smoluchowski equations
may therefore be neglected, and one attempts to solve the integro-differential
equation: $w[x\phi '(x)+2\phi(x)]=\phi(x)\int _{0}^{\infty}K(x,y)\phi(y)\text{d}y$.
For $x\ll y$, the kernel (\ref{kernel3}) may be approximated to
$K(x,y)\simeq x^{\mu }y^{\nu }-y^{\lambda }$, and one obtains: \begin{equation}
\phi (x)=x^{-\tau }\exp \left[\frac{x^{\mu }p_{\nu }}{w\mu }\right]\end{equation}
 where $p_{i}$ is the $i^{th}$ moment of the rescaled distribution
$\phi (x)$, and the exponent $\tau $ is given by $\tau =2+p_{\lambda }w^{-1}$.
It is clear that $\lim _{x\rightarrow 0}[\phi (x)]=0$. As $x$ increases
from zero, $\phi (x)$ also increases, until reaching a maximum at
$x_{m}=({w\tau }/{p_{\nu }})^{1/{\mu }}$. For $x_{m}\ll x\ll 1$
the distribution has an approximately algebraic decay $\phi (x)\sim x^{-\tau }$.
This `bell-shaped' curve is consistent with the behaviour seen in
the computer simulations when $\alpha <\frac{1}{2}$.

In the case $d_{f}>\alpha ^{-1}$, it has been shown \cite{vdande85}
that for all kernels with $\mu =0$, $\nu \leq 1$ the cluster size
distribution diverges as $x\rightarrow 0$ with the form \begin{equation}
\phi (x)\sim x^{-\tau }\end{equation}
 where $\tau =2-p_{\lambda }w^{-1}$. This behaviour is consistent
with the simulation for $\alpha \geq \frac{1}{2}$. The change in
the qualitative shape of $\phi (x\ll 1)$ around $\alpha =\frac{1}{2}$
then is further evidence to suggest that the system selects to sit
at $\nu =1$.

The shape of $\phi (x\gg 1)$ has also been studied by Van Dongen
and Ernst \cite{vdande87}. They have shown that for non-gelling kernels,
the tail of the distribution is expected to take the form \begin{equation}
\phi (x)\sim x^{-\theta }\text {e}^{-\delta x}\label{tail}\end{equation}
 where $\theta $ and $\delta $ are constants. This would appear
to be consistent with the behaviour observed in the simulations for
all values of $\alpha $, providing an exponentially dominated cut-off
at large $x$.

\section{Application to snowflake formation}

The principle motivation for the model presented in this paper was
to attempt to understand some of the properties of Cirrus clouds.  These are high altitude clouds with a base betwen 5,500 and 14,000 metres and they are usually composed solely of ice crystals \cite{PandKlett}. Amongst others, Heymsfield and Platt \cite{heymandplatt} have observed
that the ice crystals in these clouds are predominantly composed of columns, bullets,
bullet-rosettes and aggregates of these crystal types. It is these
aggregates which we hope to model, since the dominant mechanism by
which they grow is believed to be through differential sedimentation
(eg. Field and Heymsfield \cite{field}). We therefore ignore the effects of diffusional
growth, turbulence, mixing, and particle breakup, in order to concentrate
on the effects of this mechanism alone. The Reynolds number for aggregates
of a few crystals is typically between $\simeq 10$---$100$ which
ought to be modelled acceptably by our inertial flow approximation.
Because we have not modelled the detailed hydrodynamics we may also
be ignoring subtleties such as wake capture.

All of the results below are presented for the purely inertial regime
(assumed to be the most relevant to this problem) where $\alpha =\frac{1}{2}$.
The initial particles were rods of zero thickness - however, the asymptotic
behaviour is anticipated to be insensitive to the initial conditions,
and indeed by running the simulation with `bullet rosettes' for the
initial particles (three rods, crossing one another at right angles,
through a common centre), no change in the end results were found,
only in the approach to scaling.

\begin{figure}
\includegraphics{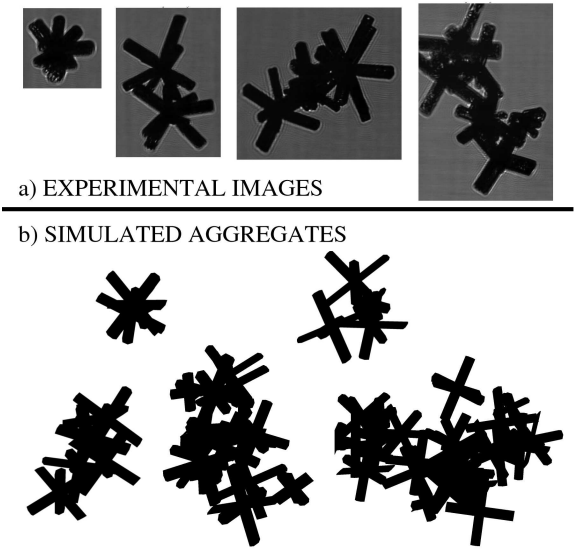}
\caption{\label{images} Projected images of a) ice crystal aggregates obtained using a cloud particle imager (CPI, SPEC Inc., USA) during an aircraft flight through a Cirrus cloud at temperatures between $-44^{\circ}$C to $-47^{\circ}$C, and b) sample clusters from our simulations.}
\end{figure}

Ice crystal aggregates have been studied through the use of cloud particle imagers during aircraft flights through ice clouds. Sample images from such a flight are shown in figure \ref{images}, alongside some of our simulation clusters. Using this experimental data, the geometry and size distribution of ice particles in these clouds has been studied, allowing for quantitative comparison between theory and experiment.

The fractal dimension of snowflakes in Cirrus may be inferred
from the work of Heymsfield \textit{et al} \cite{heyma}. By measuring
the effective density $\rho _{e}$ of bullet and bullet-rosette aggregates
as a function of their maximum linear dimension $D$, and fitting
a power law to their data, they found the relationship $\rho _{e}\sim D^{-0.96}$.
This scaling implies that the aggregates have a fractal dimension
of $d_{f}=2.04$, which is consistent with the values
predicted by our model (simulation giving $d_{f}=2.05\pm 0.1$ and
theory giving $d_{f}=2$).

\begin{figure}
\includegraphics{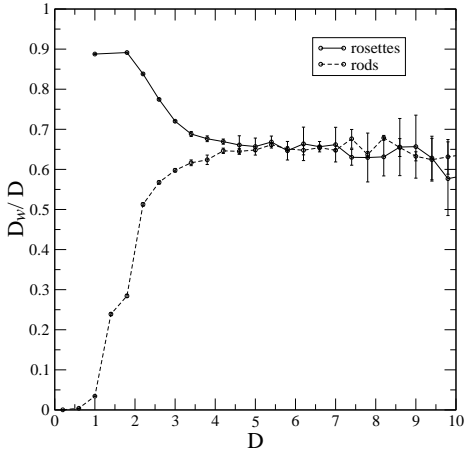}
\caption{\label{aspect}Aspect ratio of simulation clusters as a function of their maximum dimension. The
curve seems to approach an asymptotic value of $\simeq 0.65$, independent of the initial conditions used: here we show data for both rods and rosettes.}
\end{figure}

The aspect ratio of the clusters may also be calculated. Random projections of simulation clusters were taken. The
maximum dimension of the projection $D$ was measured, as was the
maximum dimension in the direction perpendicular to that longest axis,
$D_{w}$. The ratio of these two spans were binned by maximum dimension, averaged, and plotted
as a function of $D$ as shown in figure \ref{aspect}. The ratio
quickly approaches an asymptotic value of approximately $0.65\pm 0.05$.
This compares well to the measurements of Korolev and Isaac \cite{korolevandisaac},
where the ratio seems to approach a value of $\simeq 0.6-0.7$. %

\begin{figure*}
\includegraphics{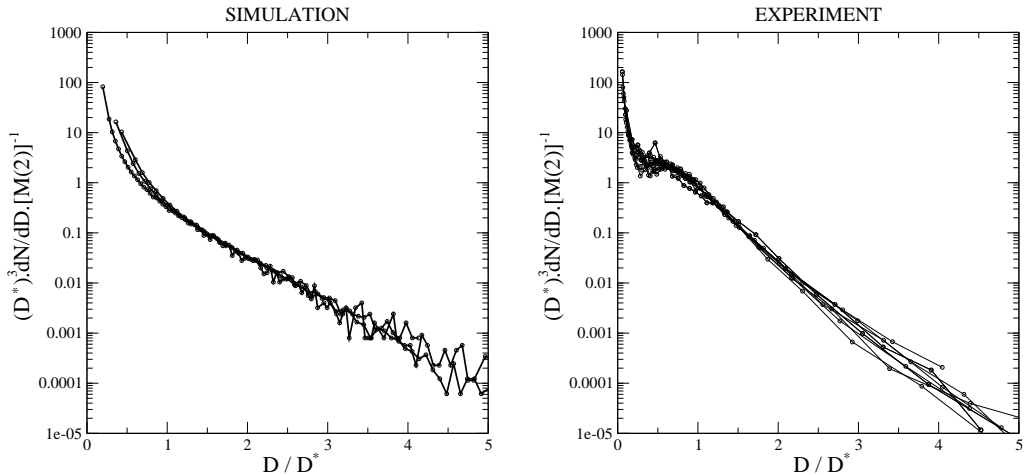}
\caption{\label{rdist}The left hand panel shows the distribution of clusters by linear size
at various stages of the simulation, rescaled in such a way as to collapse the data (see appendix). Initial conditions were 250,000 rods and the parameter $\alpha$ was set to a value of $\frac{1}{2}$. The right hand panel is a test of the same scaling using
the experimental data presented by Field and Heymsfield \cite{field}.}
\end{figure*}

Finally, the shape of the snowflake distribution of linear size may
also be compared with experiment. Field and Heymsfield \cite{field}
presented particle size distributions of the maximum length $D$ of ice particles in a Cirrus cloud. The data were obtained with an aircraft and represent in-cloud averages of particle size distributions (number per unit volume per particle size bin width) along 15km flight tracks ranging from an altitude of 9500m ($-50^{\circ}$C) to 6600m ($-28^{\circ}$C). To compare this data to the distributions
obtained from simulation, we first normalise the data, and then make
use of the dynamical scaling form (\ref{dynamicalscaling}), to collapse
the distributions onto a single curve. Details of this are given in
the appendix to this paper. The resulting histograms are shown in
figure \ref{rdist} and appear to show quite good agreement, given
the level of approximation present in our model. %

\section{Discussion and conclusions}

A simple mean-field model of aggregation by differential sedimentation
of particles in an inertial flow regime has been constructed, simulated
by computer, and analysed theoretically in terms of the Smoluchowski
equations. It has been shown that there is strong numerical evidence,
in addition to a theoretical argument, to back up the idea that the
polydispersity of the distribution and the fractal dimension feed
back on one another in such a way as to stabilise the system at $\nu =1$.
Above this value, the dominance of collisions between clusters of
very different sizes is so great as to push $d_{f}$ towards a value
of three. This in turn pulls the exponent $\nu $ back down to unity.
For $\nu <1$ the system is quite monodisperse, resulting in relatively
many collisions between clusters of similar sizes, and the fractal
dimension is reduced, forcing $\nu $ back up. The discontinuity in
the shape of the distribution around $\nu =1$ is thought to provide
the mechanism by which the system can stabilise at that point.

If it is accepted that $\nu \rightarrow 1$, then the fractal dimension
of the clusters produced may be predicted, and figure \ref{dfvsalpha}
shows that this prediction agrees well with simulation results for
$0\leq \alpha \leq \frac{2}{3}$. The sudden change in the behaviour
of $d_{f}(\alpha )$ and in the small-$x$ form of the cluster size
distribution around $\alpha =\frac{1}{2}$ is strong evidence for
the self-organisation proposed between $d_{f}$ and $\nu $.

For $\alpha >\frac{2}{3}$ the system is forced into a regime where
$\nu >1$, which has been regarded as unphysical because the Smoluchowski
equation (\ref{smol2}) predicts infinite clusters in zero time \cite{vandongen}.
In the light of our results this regime merits further study beyond
the Smoluchowski equation approximation \cite{workinprogress}.
The value $\alpha =1$ is given by viscous flow, but here our form for $\Gamma_{ij}$ does not include all of the relevant physics:
in particular, small clusters may be caught in the fluid flow, and
swept around larger clusters rather than hitting them, reducing the
dominance of big-little collisions. This has been discussed in more
detail for the particle-cluster aggregation case by Warren \textit{et
al} \cite{warrenetal}.

The application of the model to the formation of ice crystal aggregates
in Cirrus clouds has been considered: the fractal dimension, aspect
ratio, and shape of the cluster size distribution seen in the model
were all found to be consistent with experimental data. This is a
promising indication that the ideas presented in this paper may be
an acceptable model for the essential physics of snowflake aggregation
in Cirrus.

\begin{acknowledgments}
We are grateful to Aaron Bansemer for initial processing of the size distribution data and Carl Schmitt for supplying the CPI images.
This work was supported financially by the Engineering and Physical
Sciences Research Council, and The Meteorological Office.
\end{acknowledgments}
\appendix*

\section{Scaling of the cluster radius distribution}

Experiments have reported the distribution of ice aggregates by linear
span rather than by mass, and we present here how that distribution
$\frac{\text {d}N}{\text {d}D}$ should naturally be rescaled. This tests the
dynamical scaling ansatz which, for the mass distribution, gave $\frac{\text {d}N}{\text {d}m}=n_{m}=s^{-a}\phi (m/s)$,
where $a=2$ in mass-conserving systems. We anticipate fractal scaling
so that $m\sim D^{d_{f}}$ and hence: \begin{equation}
\frac{\text {d}N}{\text {d}D}\sim D^{d_{f}-1}s^{-a}\phi \left(\frac{m}{s}\right).\end{equation}
 From this expression we may calculate the moments of the distribution
$M(b)\equiv \int \frac{\text {d}N}{\text {d}D}D^{b}\text {d}D$ in
terms of the average cluster mass $s(t)$: \begin{equation}
M(b)\sim s^{-a+1+b/d_{f}}\int _{1/s}^{\infty }x^{b/d_{f}}\phi (x)\text {d}x\end{equation}
 where $x=m/s$. At small sizes we expect $\phi (x)\sim x^{-\tau }$.
If $b>d_{f}(\tau -1)$ therefore, the integral converges as $s\rightarrow \infty $,
and $M(b)\sim s^{-a+1+b/d_{f}}$. From our simulations, we have measured
$\tau \simeq 1.6$, $d_{f}\simeq 2$, and so the lowest integer moment
which scales in this way is the second. We therefore choose this to
normalise our data: \begin{equation}
[M(2)]^{-1}\frac{\text {d}N}{\text {d}D}\sim D^{d_{f}-1}s^{-1-2/d_{f}}\phi \left(\frac{m}{s}\right)\end{equation}
 which, defining the average cluster diameter $D^{*}\equiv M(3)/M(2)\sim s^{1/d_{f}}$
yields: \begin{equation}
[M(2)]^{-1}\frac{\text {d}N}{\text {d}D}\sim (D^{*})^{-3}\psi \left(\frac{D}{D^{*}}\right),\end{equation}
where $\psi(y)=y^{d_{f}-1}\phi(y^{d_{f}})$. Hence, if we assume
that $d_{f}$ approaches a constant value, plots of $\{[M(2)]^{-1}.\frac{\text {d}N}{\text {d}D}.(D^{*})^{3}\}$
against $(D/D^{*})$ should all lie on a single curve.

\end{document}